\documentclass[%
superscriptaddress,
preprint,
showpacs,
 amsmath,amssymb,
 aps,
 prc
]{revtex4-1}

\usepackage{graphicx}
\usepackage{dcolumn}
\usepackage{bm}
\usepackage{soul,xcolor}
\usepackage{CJK}

\setstcolor{blue}

\begin{document}
\begin{CJK*}{UTF8}{}


\title{Thermodynamics of pairing transition in hot nuclei}

\author{Lang Liu {\CJKfamily{gbsn}(刘朗)}}
 \affiliation{School of Science, Jiangnan University, Wuxi 214122, China.}
 \affiliation{State Key Laboratory of Nuclear Physics and Technology, School of Physics, Peking University, Beijing 100871, China.}

 \author{Zhen-Hua Zhang {\CJKfamily{gbsn}(张振华)}}%
\affiliation{State Key Laboratory of Nuclear Physics and Technology, School of Physics, Peking University, Beijing 100871, China.}
\affiliation{Mathematics and Physics Department, North China Electric Power University, Beijing 102206, China.}%


 \author{Peng-Wei Zhao {\CJKfamily{gbsn}(赵鹏巍)}}
 \email{pwzhao@pku.edu.cn}
 \affiliation{Yukawa Institute for Theoretical Physics,
Kyoto University, Kyoto 606-8502, Japan.}
\affiliation{State Key Laboratory of Nuclear Physics and Technology, School of Physics, Peking University, Beijing 100871, China.}



\begin{abstract}
The pairing correlations in hot nuclei $^{162}$Dy are investigated in terms of the thermodynamical properties by covariant density functional theory. 
The thermodynamical quantities are evaluated by the canonical ensemble theory and the paring correlations are treated by a shell-model-like approach, in which the particle number is conserved exactly.  An S-shaped heat capacity curve as a function of temperature has been obtained. The properties of hot nuclei, such as entropy and level density are studied in terms of defined seniority component.
It is found that the one-pair-broken states play crucial roles in the appearance of the S shape of the heat capacity curve. Moreover, due to the effect of the particle-number conservation, the pairing gap varies smoothly with the temperature, which indicates a gradual transition from the superfluid to the normal state.
\end{abstract}

\maketitle
\end{CJK*}


\section{Introduction}

A phase transition is well defined for infinite systems, while for finite many-body systems, its realization is often obscured due to the surface effects and statistical fluctuations. The ground states of most nuclei, i.e., at zero temperature, are superfluid states, but in warm nuclei~\cite{Bohr1998NuclearStructure:I} the superfluidity tends to be vanishing when the temperature increases. Such superfluid-to-normal transition has attracted wide attentions and, in the past decades, great progress has been achieved in the experimental aspects, thanks to the accurate measurements of the level density~\cite{Schiller2001PRC63:021306,Melby2001PRC63:044309,Melby1999PRL83:3150--3153,Guttormsen2003PRC68:034311}. From these investigations,  the so-called S-shaped curve of heat capacity has been found as a function of temperature. This was regarded as a fingerprint of the superfluid-to-normal (pairing) phase transition. Based on this picture, the critical temperature has been estimated from the experimental data as $T_c \simeq 0.5$ MeV for $^{161,162}$Dy, $^{171,172}$Yb~\cite{Schiller2001PRC63:021306}, and $^{166,167}$Er~\cite{Melby2001PRC63:044309}.

Similar S shapes come out in many theoretical calculations as well and the nature of the S-shaped heat capacity has also been discussed for many years in the framework of shell model~\cite{Rombouts1998PRC58:3295--3304,Liu2001PRL87:022501,Langanke2005NPA757:360--372}, mean field models~\cite{Egido2000PRL85:26--29,Agrawal2000PRC62:044307,Sandulescu2000PRC61:044317,Niu2013PRC88:034308,Li2015PRC92:014302} and other models see, e.g., Refs.~\cite{Guttormsen2001PRC63:044301,Guttormsen2001PRC64:034319}. In particular, within the mean-field picture, one could define the superfluid and the normal-fluid phases in nuclei with Bardeen-Cooper-Schrieffer (BCS) theory or the Hartree-Fock-Bogoliubov theory. Clear signatures of pairing phase transition have been provided by the fact that the S shapes of heat capacity could be reproduced by most mean-field calculations including finite-temperature BCS~\cite{Sano1963PTP29:397--414,Goodman1981NPA352:30--44,Gambacurta2013PRC88:034324}, finite-temperature HFB with a pairing-plus-quadrupole Hamiltionian~\cite{Goodman1986PRC34:1942--1949}, as well as the self-consistent mean-field models in both nonrelativistic~\cite{Reiss1999EPJA6:157--165} and relativistic form~\cite{Niu2013PRC88:034308,Li2015PRC92:014302}. With a variety of quantum fluctuations, it has been found that the critical temperatures for the pairing phase transition given by most mean-field models locate in the interval of $0.5$--$0.6 \Delta(0)$, where $\Delta(0)$ is the pairing energy gap at zero temperature.

Breakdown of a certain symmetry is often associated with phase transitions. For the pairing phase transition, within the mean-field theories, the particle number conservation is violated in the superfluid phase while preserved in the normal-fluid phase. The number conservation effects on the nuclear heat capacity has been investigated through the particle-number projection methods based on the finite-temperature BCS or HFB approaches~\cite{Esebbag1993NPA552:205--231,Esashika2005PRC72:044303,Gambacurta2013PRC88:034324}. Due to the restoration of particle number conservation, the calculated heat capacity varies smoothly with the temperature, indicating a gradual transition from the superfluid to the normal phase.

Therefore, it is imperative to investigate the pairing transition in hot nuclei and the nature of the S-shaped heat capacity curve by the shell-model-like approach (SLAP)~\cite{Zeng1983NPA405:1--28,Zeng1994PRC50:1388--1397,Meng2006FPC1:38--46}. In this approach, the particle number is strictly conserved and the blocking effects are also treated exactly. On the other hand, due to the successful description of many nuclear phenomena, covariant density functional theory CDFT has been one of the most important microscopic models for nuclear structure~\cite{Ring1996PPNP37:193--263,Meng2006PPNP57:470--563,Vretenar2005PR409:101--259}. The CDFT includes the complicated interplay between the large Lorentz scalar and vector self-energies induced on the QCD level, and naturally treats the spin degrees of freedom. The shell-model-like approach has been implemented in the framework of CDFT in Ref.~\cite{Meng2006FPC1:38--46} in 2006. In this work, the shell-model-like approach will be applied to the investigation of the S shape of the heat capacity in hot nuclei $^{162}$Dy in the framework of CDFT.

\section{Theoretical framework}
\label{sec:theory}

The CDFT starts from a Lagrangian and the corresponding Kohn-Sham equations have the form of a Dirac equation for nucleons with effective fields $S(\bm{r})$ and $V(\bm{r})$ derived from this Lagrangian
\begin{eqnarray}
   \{ \bm{\alpha}\cdot\bm{p}+V(\bm{r})+\beta[M+S(\bm{r}) ] \}\psi_i = \varepsilon_i \psi_i,
   \label{DiracE}
\end{eqnarray}
where $\varepsilon_i$ is the single particle energy of the Dirac state $i$. The scalar $S(\bm{r})$ and vector $V(\bm{r})$ potentials are connected in a self-consistent way to various densities through the Klein-Gordon equations for the meson fields $\sigma(\bm{r})$, $\omega(\bm{r})$, and $\rho(\bm{r})$ and the photon fields $A(\bm{r})$,
\begin{eqnarray}
\left\{
\begin{array}{l}
\left[ -\triangle+m^2_{\sigma} \right] \sigma(\bm{r}) = -g_{\sigma}\rho_{s}(\bm{r})-g_2\sigma^2(\bm{r})-g_3\sigma^3(\bm{r}) , \\
\left[ -\triangle+m^2_{\omega} \right] \omega(\bm{r}) = g_{\omega}\rho_v(\bm{r})-c_3\omega^3(\bm{r}), \\
\left[ -\triangle+m^2_{\rho} \right] \rho(\bm{r}) = g_{\rho}\rho_3(\bm{r}), \\
 -\triangle\,A(\bm{r}) = e\rho_{p}(\bm{r}).
\end{array}
\right.
\label{KGE}
\end{eqnarray}
Following the definition of the Dirac spinors in Ref.~\cite{Meng2006FPC1:38--46}, the densities can be represented as
\begin{eqnarray}
\rho_{s,v} = 2\sum\limits_{i} \left[ \left( |f^+_i|^2+|f^-_i|^2 \right) \mp \left( |g^+_i|^2+|g^-_i|^2 \right) \right],
\label{den1}
\end{eqnarray}
where $f_i$ and $g_i$ represent, respectively, the large and small components of the Dirac state $i$.

The iterative solution of these equations yields the total energy, quadrupole moments, single-particle energies, etc. 

The total energy calculated by the RMF for the system is:
\begin{equation}
E_{\rm RMF}=E_{\rm nucleon}+E_\sigma+E_\omega+E_\rho+E_{\rm c}+E_{\rm CM},
\end{equation}
with
\begin{eqnarray}
\left\{
\begin{array}{lll}
E_{\rm nucleon}&=&\sum\limits_{i}{\varepsilon_i}\\
E_\sigma&=&-\frac{1}{2}\int{d}^{3}r
\left\{
g_{\sigma}\rho_{\rm s}({\bf r})\sigma(r)+[\frac{1}{3}g_2\sigma({\bf r})^3
+\frac{1}{2}g_3{\sigma({\bf r})}^4]
\right\}\\
E_\omega&=&-\frac{1}{2}\int{d}^{3}r{\{}g_{\omega}\rho_{\rm v}({\bf r})\omega^{0}({\bf r})-\frac{1}{2}g_{4}\omega^{0}({\bf r})^{4}\}\\
E_\rho&=&-\frac{1}{2}\int{d}^{3}r{g_{\rho}\rho_{3}({\bf r})\rho^{00}({\bf r})}\\
E_{\rm c}&=&-\frac{e^2}{8\pi}\int{d}^{3}r\rho_{\rm c}(r)A^{0}({\bf r})\\
E_{\rm CM}&=&-\frac{3}{4}41A^{-1/3}\\
\end{array}
\right.
\end{eqnarray}
where $E_{\rm nucleon}$ is the summation of the energies of nucleon $\varepsilon_i$; $E_\sigma$, $E_\omega$, $E_\rho$ and $E_{\rm c}$ are the energies of the meson fields and the Coulomb fields, $E_{\rm CM}$ is the correction for the center-of-mass motion.

For open shell nuclei, one needs to take into account the pairing correlations.
In the present work, the SLAP is implemented in the framework of CDFT to treat the pairing correlations. The total Hamiltonian of SLAP reads
\begin{eqnarray}
  \label{eq:h}
  H &=& H_{\rm s.p.} + H_{\rm pair} \cr
    &=& \sum\limits_{i}\varepsilon_{i}a^{\dagger}_{i}a_{i}
  -G\sum\limits^{i\neq j}_{i,j>0} a^{\dagger}_{i}a^{\dagger}_{\bar{i}}a_{\bar{j}}a_{j},
\end{eqnarray}
where $\varepsilon_{i}$ is the single-particle energy obtained from the Dirac equation (\ref{DiracE}), $\bar i$ is the time-reversal state of $i$, and $G$ represents a constant pairing strength.
This Hamiltonian is diagonalized in a space constructed with a set of multi-particle configurations (MPCs). For a system with an even particle number $N = 2n$, the MPCs could be constructed as follows:
\begin{enumerate}
\item fully paired configurations (seniority $s = 0$):
\begin{eqnarray}
|c_1\bar{c}_1\cdots c_n\bar{c}_n\rangle=a^\dagger_{c_1}a^\dagger_{\bar{c}_1}\cdots a^\dagger_{c_n}a^\dagger_{\bar{c}_n}|0\rangle;
\end{eqnarray}
\item configurations with two unpaired particles (seniority $s = 2$)
\begin{eqnarray}
|i\bar{j}c_1\bar{c}_1\cdots c_{n-1}\bar{c}_{n-1}\rangle=a^\dagger_{i}a^\dagger_{\bar{j}}a^\dagger_{c_1}a^\dagger_{\bar{c}_1}\cdots a^\dagger_{c_{n-1}}a^\dagger_{\bar{c}_{n-1}}|0\rangle\quad\quad(i\ne j);
\end{eqnarray}
\item configurations with more unpaired particles (seniority $s=4, 6, \ldots$), see, e.g., Refs.~\cite{Zeng1983NPA405:1--28,Meng2006FPC1:38--46}. 
\end{enumerate}

The Hamiltonian (\ref{eq:h}) have the good quantum numbers of the parity $\pi$ and the seniority $s$. As a result,  the MPC space could be written as:
\begin{eqnarray}
{\rm MPC~space} &=& (s=0,\pi=+) \oplus (s=0, \pi=-) \oplus \nonumber \\
                 && (s=2,\pi=+)  \oplus  (s=2, \pi=-) \oplus \nonumber \\ 
                 && \cdots
\end{eqnarray}
In the practical calculations, the MPC space has to be truncated with an energy cutoff $E_c$, i.e., the configurations with energies $E_m-E_0 \leq E_c$ are used to diagonalize the Hamiltonian~(\ref{eq:h}), where $E_m$ and $E_0$ are the energies of the $m$th configuration and the ground-state configuration, respectively. 

After the diagonalization of the Hamiltonian (\ref{eq:h}), one could obtain the nuclear many-body wave function
\begin{eqnarray}
|\psi_\beta\rangle&=&\sum\limits_{c_{1}\cdots c_{n}}{v_{\beta,\,c_1\cdots c_n}}|c_1\bar{c}_1\cdots c_n\bar{c}_n\rangle  \cr
&& +\sum\limits_{i,j}{\sum\limits_{c_{1}\cdots c_{n-1}}{v_{\beta(ij),\,c_1\cdots c_{n-1}}}|i\bar{j}c_1\bar{c}_1\cdot\cdot\cdot c_{n-1}\bar{c}_{n-1}\rangle} \cr
&& + \cdots,
\end{eqnarray}
where $\beta = 0$ for the ground state, and $\beta = 1, 2, 3, \ldots$ for the excited states. $v_{\beta}$ means the coefficient after diagonalization.
The pairing energy then can be calculated by
\begin{eqnarray}
E_{\rm pair}&=&\langle\psi_{\beta}\mid{H_{\rm pair}}\mid\psi_{\beta}\rangle,
\end{eqnarray}
The total energy of nuclei is written as
\begin{eqnarray}
E_{\rm total} = E_{\rm RMF} + E_{\rm pair}.
\end{eqnarray}
Furthermore, the pairing gap energy could be evaluated as~\cite{Canto1985PLB161:21--26,Egido1985PLB154:1--5,Shimizu1989RMP61:131--168}
\begin{eqnarray}\label{eq:gap1}
\Delta_{\beta}=G\left[ -\frac{1}{G} \langle \Psi_{\beta} | H_{\rm p} | \Psi_{\beta} \rangle \right]^{1/2}.
\end{eqnarray}
We briefly summarize our framework as follows: We first solve the Dirac equation and the Klein-Gordon equations self-consistently. The obtained single-particles states are used to construct a many particle configuration space, which is used to diagonalize the SLAP Hamiltonian in a shell-model-like way. The expectation value of the pairing Hamiltonian is calculated with the obtained SLAP wave function; this results in the pairing energy $E_{\rm pair}$, which together with the RMF energy $E_{\rm RMF}$, gives rise to the total energy. More details of RMF+SLAP can be found in Ref.~\cite{Meng2006FPC1:38--46}.

The thermodynamic properties of the pairing interaction
are calculated here in the canonical ensemble~\cite{Sumaryada2007PRC76:024319}, whose canonical partition function $Z$, average energy $\langle E \rangle$, heat capacity $C_V$, and entropy $S$ are defined with the following equations,
\begin{eqnarray}
\label{eq:z}
Z &=& \sum\limits^{\infty}_{\beta=0}\eta(E_{\beta})\,e^{-E_{\beta}/T}, \\
\label{eq:e}
\langle E \rangle &=& Z^{-1}\sum\limits^{\infty}_{\beta=0}E_{\beta} \,\eta(E_{\beta})\,e^{-E_{\beta}/T}, \\
\label{eq:cv}
C_V &=& \frac{\partial \langle E \rangle}{\partial T},  \\
\label{eq:s}
S &=& \frac{\partial\,(T\ln Z)}{\partial T} = \frac{\langle E(T) \rangle}{T} + \ln Z,
\end{eqnarray}
where $E_{\beta}$ is the excitation energy which could be obtained from the SLAP method with CDFT, and the corresponding level density $\eta (E_{\beta})$ is taken as $2^s$, i.e., the degeneracy of each state. By means of the partition function, one can also evaluate the ensemble average pairing gap energy as
\begin{eqnarray}\label{eq:gap}
\tilde{\Delta} = Z^{-1}\sum\limits^{\infty}_{\beta=0} \Delta_{\beta} \,\eta(E_{\beta})\,e^{-E_{\beta}/T}.
\end{eqnarray}

\section{Numerical details}

In this work, the axial symmetry is imposed, and the Dirac equation~(\ref{DiracE}) is solved in a space of axially deformed harmonic oscillator basis with $14$ major shells~\cite{Ring1997CPC105:77--97}. The effective interaction PK1 is adopted~\cite{Long2004PRC69:034319}. In the construction of the multi-particle configurations, 20 single particle levels around Fermi surfaces and six pairs of valence particles, which mean the particles which are used to build the MPCs, are included for both neutrons and protons. This also indicates the highest seniority taken into account in the calculations is 12.
The effective pairing strengths can, in principle, be determined by the odd-even differences in the nuclear binding energies, and are connected with the dimension of the truncated MPC space. The odd-even mass difference is defined, e.g., for neutron, as:
\begin{eqnarray}
\Delta_n = \dfrac{1}{2}[ B(N-1,Z)+B(N+1,Z) ]-B(N,Z),
\end{eqnarray}
where the $B(N,Z)$ is the binding energy of the nucleus with neutron number $N$ and proton number $Z$. The pairing strength in our calculations for neutron $G_{\rm n}$ is fixed to 0.29 MeV, and $G_{\rm p}$ for proton is 0.32 MeV with $E_c$=30 MeV. Note that more sophisticated calculations with, for example, angular momentum projection may lead to slight changes for the calculated odd-even mass differences.

\begin{figure}[ht!]
\centering
 \includegraphics[scale=0.8]{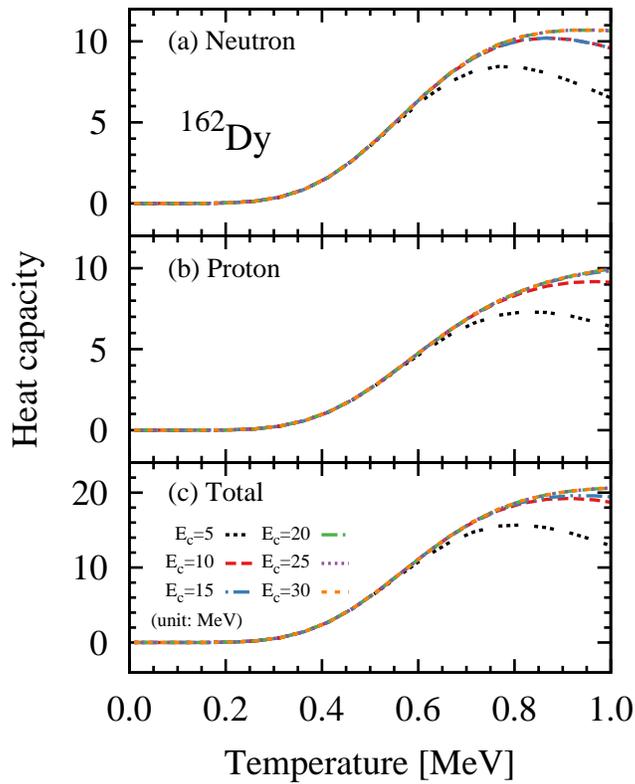}
 \caption{(Color online) Neutron (a), proton (b), and the total (c) heat capacities for $^{162}$Dy calculated with $E_c$=5, 10, 15 20, 25, and 30 MeV as functions of the temperature.}
\label{fig:ec}
\end{figure}

First of all, the calculated results with increasing energy cut off $E_c$ are checked. Figure~\ref{fig:ec} shows the neutron (a), proton (b), and the total (c) heat capacities for $^{162}$Dy calculated with $E_c$=5, 10, 15 20, 25, and 30 MeV as functions of the temperature. It can be found that the calculated heat capacities with $E_c = 20, 25$, and 30 MeV obtain consistent values from 0 to 1 MeV. In the following calculations, the energy cutoff $E_c$ is fixed as 30 MeV. The dimensions of the proton and neutron MPC spaces are about $3\times10^5$ and $5\times10^5$, respectively.

\section{Results and discussion}

\begin{figure}[ht!]
\centering
 \includegraphics[scale=.8]{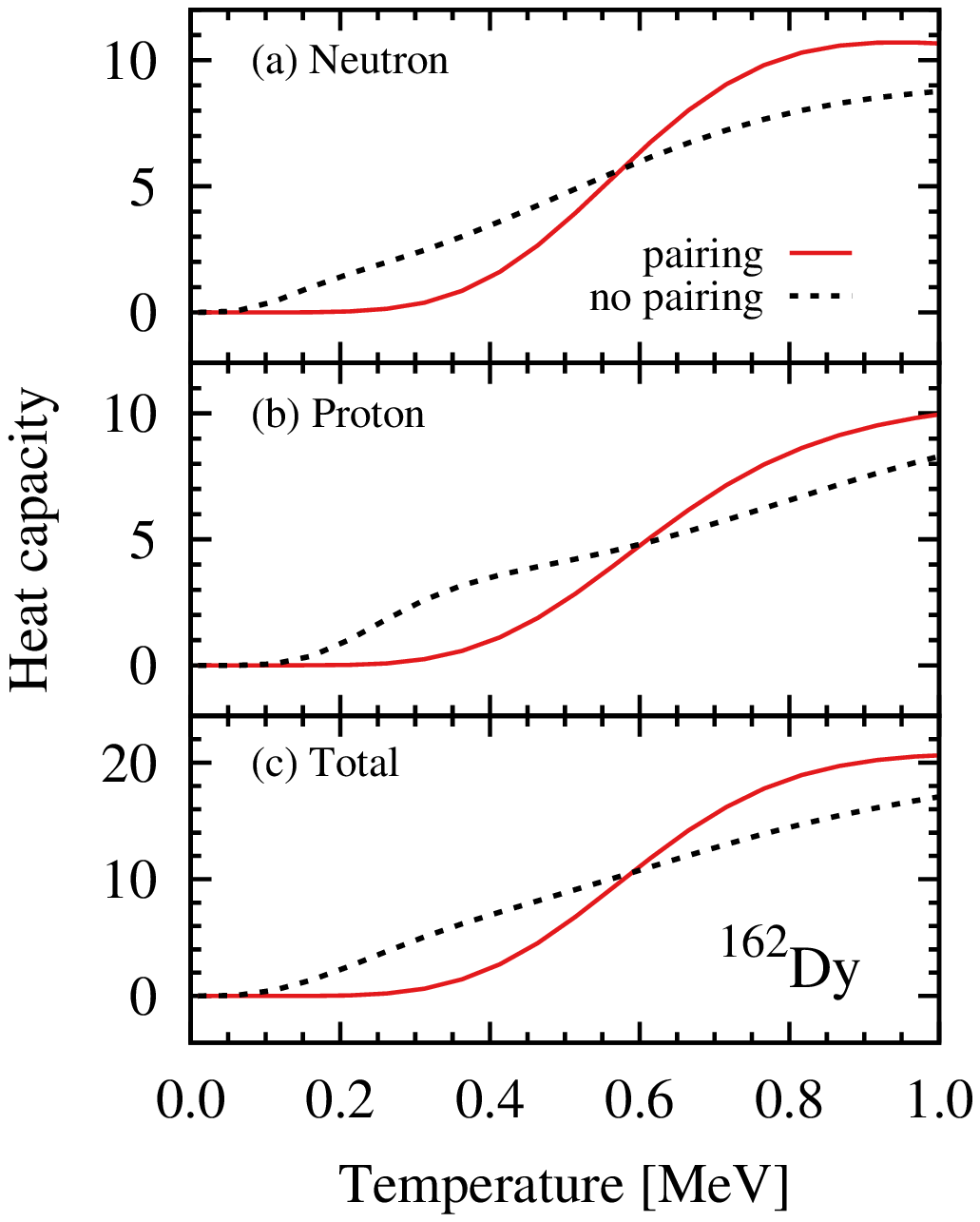}
 \caption{(Color online). Neutron (a), proton (b), and the total (c) heat capacities for $^{162}$Dy as functions of the temperature with (red solid lines) and without pairing (black dashed lines). }
\label{fig:cv}
\end{figure}

The heat capacity can be evaluated from the partial derivative of the average energy with respect to the temperature as expressed in Eq.~(\ref{eq:cv}). Because the proton and neutron degrees of freedom are treated separately in the Hamiltonian of Eq.~(\ref{eq:h}), the heat capacity could be straightforwardly divided into two parts which correspond to the proton and neutron excitations, respectively. In Fig.~\ref{fig:cv}, the neutron, proton as well as the total heat capacities for $^{162}$Dy are shown as functions of temperature. For comparison, the results calculated without pairing correlations ($G=0$ MeV) are shown as dashed lines. 
It can be seen that the heat capacities without pairing increase almost linearly with the temperature. This linear tendency is very analogous to the results of a pure Fermi gas model~\cite{Schiller2001PRC63:021306}, while the gentle fluctuations shown in the proton heat capacity result from the shell structures in the single proton levels of CDFT.
The observed S shape in the experimental heat capacity is reproduced from our calculations when the pairing correlations are taken into account. 
The calculated heat capacities are nearly zero at low temperature ($T\le 0.35$ MeV) due to the large energy gap between the ground state and two-quasi-particle excited states induced by the inclusion of the pairing correlations. 
When the temperature grows up, many pair-broken excited states with seniority $s=2, 4, ...$ appear, and thus the heat capacity increases rapidly till the inflection point at $T\approx 0.75$ MeV.  
Above this point, the heat capacity increases much slower with the temperature.



\begin{figure}[ht!]
\centering
 \includegraphics[scale=.8]{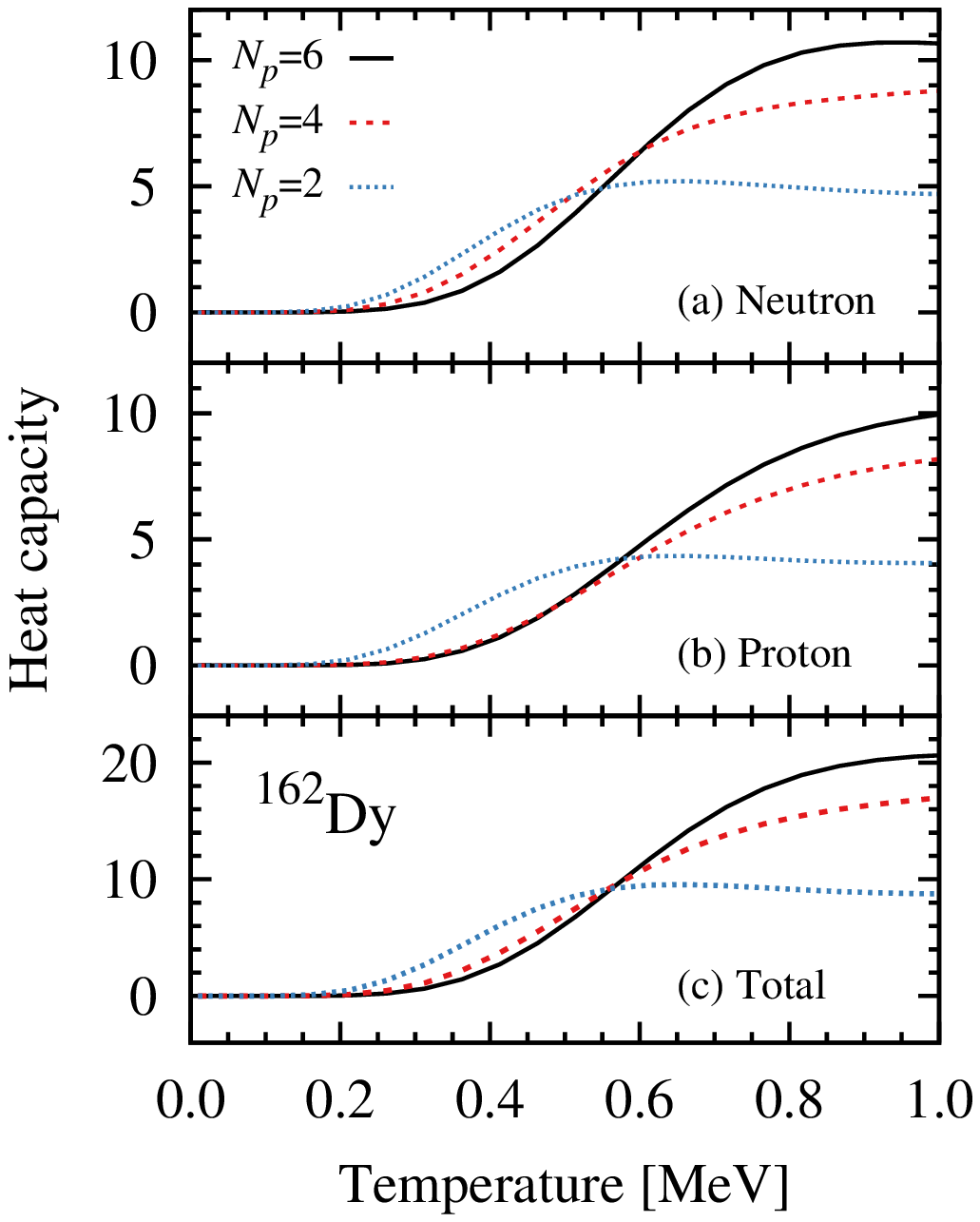}
 \caption{(Color online) Neutron (a), proton (b), and the total (c) heat capacities for $^{162}$Dy calculated with different number of valence particle pairs $N_p$=2 (blue dotted lines), 4 (red dashed lines) and 6 (black solid lines) in the model space as functions of the temperature.}
\label{fig:cv_pair}
\end{figure}

It should be mentioned that the particle number in the model space may influence the behavior of the heat capacity at very high temperature. This could be clearly seen in Fig.~\ref{fig:cv_pair}, where the heat capacities calculated with different number of valence particle pairs ($N_p = 2,4,6$) in the model space are shown as functions of the temperature. 
The heat capacity curves are almost identical for $N_p =  2,4,6$ below $T\sim 0.2$ MeV since almost no pair break happens at such low temperature.
From there on, these heat capacity curves start to deviate from each other.  
In particular, the heat capacity for $N_p = 2$ drops at $T \sim 0.6$ MeV. 
This is due to the fact that the two valence proton and neutron pairs in the model space are exhausted and not able to continue to absorb energy with the same rate. Moreover, it is found that the values of the heat capacity obtained in this case are roughly $50\%$ lower than the observed ones.
It has been known that if too few pairs are contained, one may easily misinterpret the S shape of the heat capacity curve and underestimate the value of the heat capacity~\cite{Guttormsen2001PRC63:044301}. 
In order to avoid such a misinterpretation of the S shape here, more Cooper pairs are included in the model space. 
It turns out here that for $^{162}$Dy one could get reasonable values of the heat capacity after six valence proton and neutron pairs are taken into account considering the calculation resources. This is also consistent with the previous work as in Ref.~\cite{Guttormsen2001PRC63:044301}.



\begin{figure}[ht!]
  \centering
    \includegraphics[scale=.8]{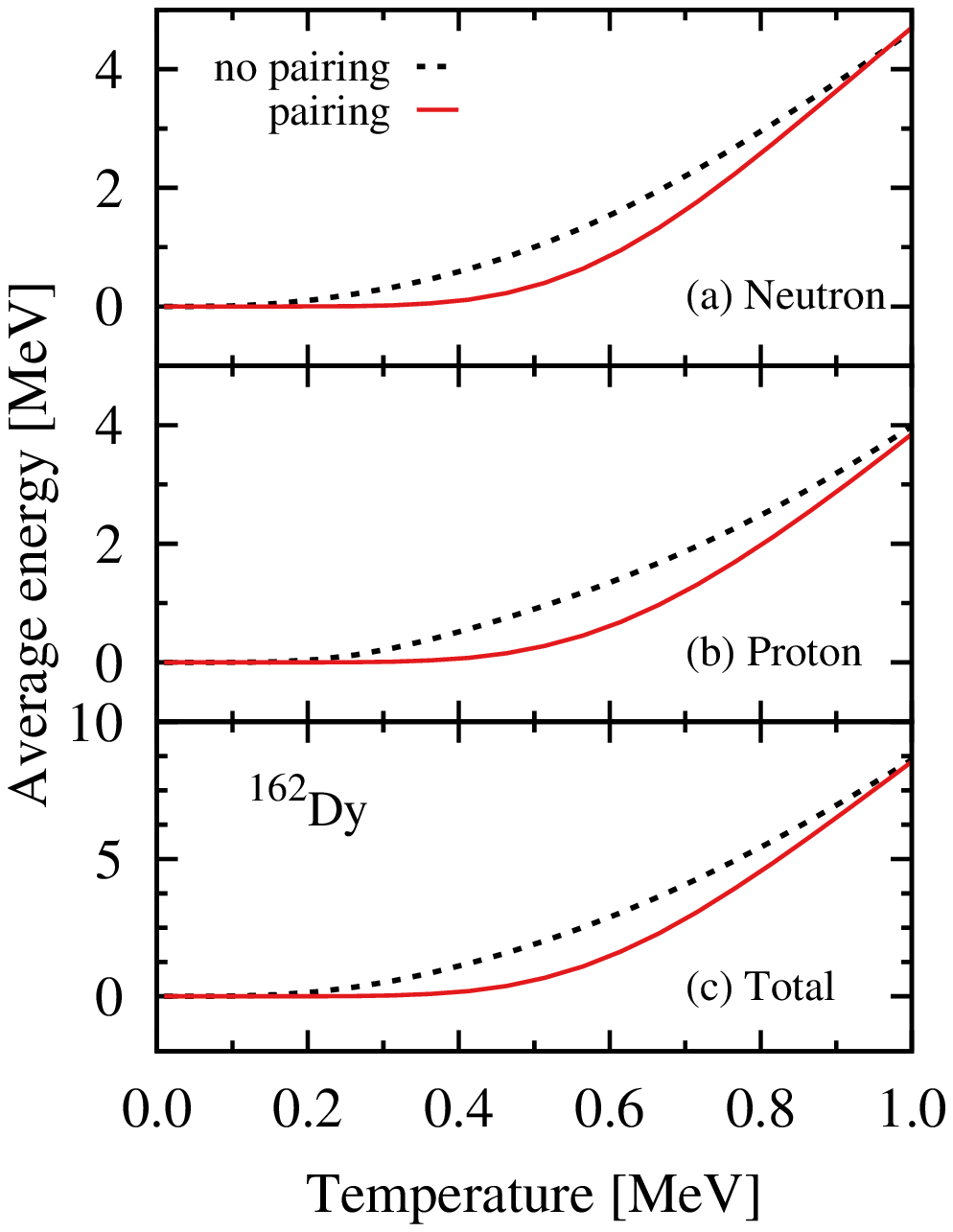}
  \caption{(Color online) Neutron (a), proton (b), and the total (c) average excitation energies for $^{162}$Dy calculated with (red solid line) and without (black dashed line) pairing correlations as functions of the temperature.}
\label{fig:e}
\end{figure}

Usually, the thermodynamic properties of hot nuclei are described at the temperature representation in this work. At the same time, it is also convenient to study those hot nuclei in terms of excitation energies. The neutron (a), proton (b), and the total (c) average excitation energies defined in Eq.~(\ref{eq:e}) for $^{162}$Dy calculated with (red solid line) and without (black dashed line) pairing correlations as functions of the temperature are shown in Fig.~\ref{fig:e}, which exhibits directly the correspondence between the temperature and the average excitation energy. It can be seen that, when pairing is considered, the  average excitation energies for both neutrons and protons is about zero below $T\sim$ 0.4 MeV. While around $T~\sim 0.9$ MeV the average excitation energies are comparable to the binding energy per nucleon. The calculated heat capacities without pairing are also shown in the same figure for comparison. It can be seen that the heat capacity curves without pairing is almost linear increasing beyond $T\sim$ 0.2 MeV. In addition, the system with pairing requires higher temperature to obtain the same excitation energy than those without pairing.



\begin{figure}[ht!]
  \centering
    \includegraphics[scale=.8]{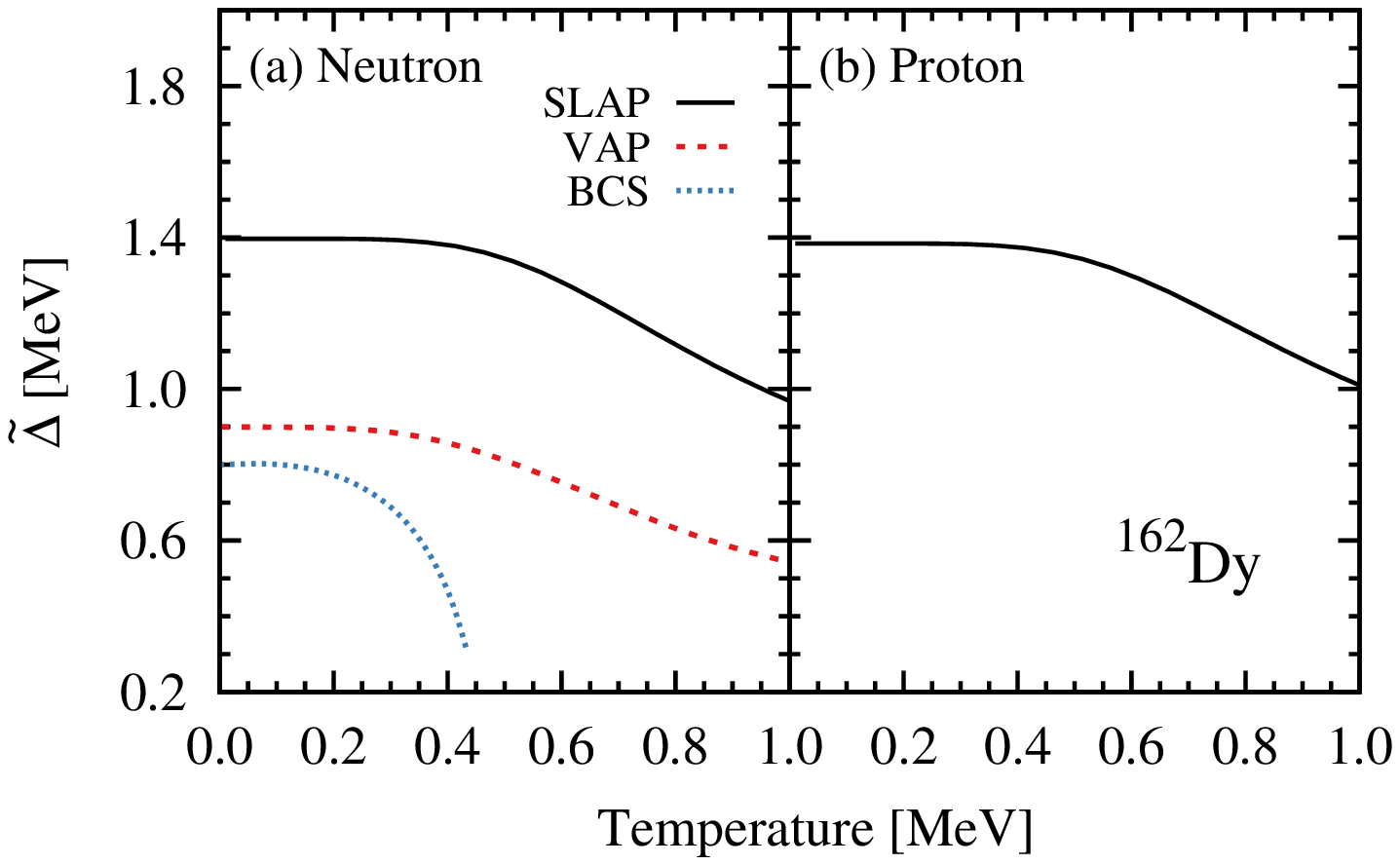}
  \caption{(Color online) Neutron (a) and proton (b) pairing gap energies for $^{162}$Dy as functions of the temperature calculated by SLAP (black solid lines). For comparison, the neutron pairing gap energies obtained from finite temperature BCS (blue dotted line) and finite temperature variation after projection (VAP) BCS (red dashed line) approaches~\cite{Gambacurta2013PRC88:034324} are also shown.}
\label{fig:gap}
\end{figure}

 The transition of the pairing correlations with the temperature is characterized by the pairing gap energies. 
Figure~\ref{fig:gap} shows how the neutron and proton pairing gaps vary with the temperature.  
The fact that the pairing gap is almost constant below $T \sim 0.35$ MeV is connected with the nearly zero heat capacity as shown in Fig.~\ref{fig:cv}.
Above $T \sim 0.35$ MeV, the neutron pairing gap calculated by SLAP decreases smoothly with the temperature, while it does not vanish at high temperature up to $1$ MeV.
This indicates a gradual pairing transition from the superfluid state to the normal state in the hot nucleus $^{162}$Dy.
The drop of the pairing gap results in an increasing number of the Cooper-pair-broken excited states, and thus a rapid increase of the heat capacity.
In this way, the S shape of heat capacity curves shown in Fig.~\ref{fig:cv} is provided by the competition between the effects from temperature and pairing correlations. 

For comparison, the neutron pairing gap energies calculated in the finite-temperature BCS and variation after projection BCS approaches~\cite{Gambacurta2013PRC88:034324}, are also shown in Fig.~\ref{fig:gap}. These results are systematically smaller than those obtained from SLAP, since as in Ref.~\cite{Gambacurta2013PRC88:034324}, the pairing strength $G$ there is fixed to have a pairing BCS gap of 0.8 MeV. Nevertheless, it is not the magnitude but its variation tendency with the temperature is the main focus here. 
One could see that the finite-temperature BCS predicts a sharp transition from the superfluid to the normal phase. 
As it is well known, this sharp transition is connected to the particle number violation. Due to the restoration of particle number conservation, the pairing gap calculated in the VAP approach varies smoothly with the temperature, which is very similar to the present SLAP results. It should be noted that the present SLAP calculation has provided an exact treatment of pairing correlations without any particle number fluctuations, and it could be implemented easily and effectively to the self-consistent framework of the density functional theory~\cite{Meng2006FPC1:38--46}. The extension of such model and its application in the investigation of the nuclear pairing in hot nuclei are in progress.    



\begin{figure}[ht!]
\centering
 \includegraphics[scale=.8]{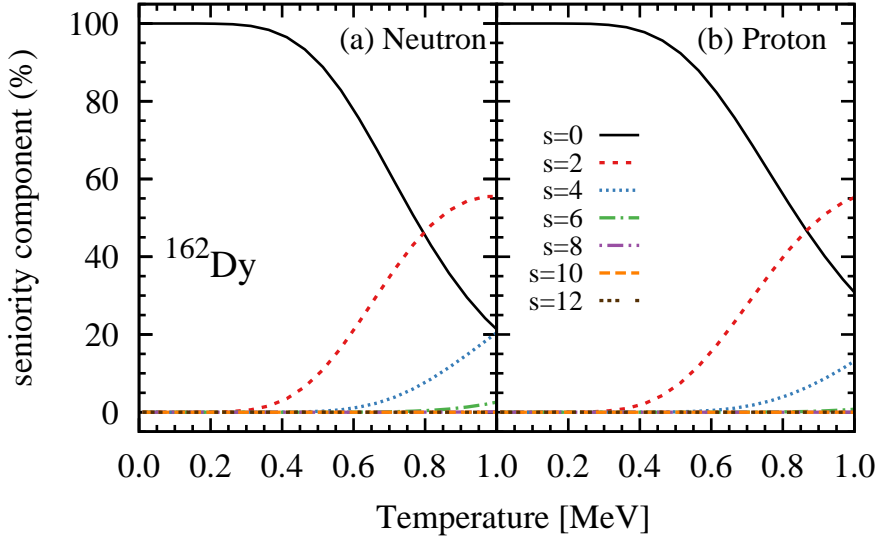}
 \caption{(Color online) Neutron (a) and proton (b) seniority components $\chi_s$ for $^{162}$Dy with different seniority numbers $s = 0, 2, 4, 6, 8, 10, 12$ as functions of the temperature. }
 \label{fig:probability}
\end{figure}

In order to provide a microscopic picture of the nuclear pairing transition, it is interesting to explore how many Cooper pairs would be broken with the increasing temperature in the nucleus $^{162}$Dy,. Here, we define the so-called \emph{seniority component} 
\begin{equation}
  \chi_s = Z^{-1} \sum\limits_{\beta\in \{s\}} \eta(E_{\beta}) e^{-E_{\beta}/T},
\end{equation}\label{s_com} 
where the summation runs over only the excited states with their seniority number $s = 0, 2, 4, ...$. With such definition, the average energy could be rewritten as $\langle E \rangle = \sum\limits_{s,\beta}\chi_s E_{\beta\in s}$. Here, it is very clear that the seniority component here just represents the contribution of the excited states with each seniority number to the total average energy.
In Fig.~\ref{fig:probability}, the neutron and proton seniority components $\chi_s$ with different seniority numbers $s = 0, 2, 4, 6, 8, 10, 12$ are shown as functions of the temperature, respectively.
One could see that the $s=0$ states contribute almost $100\%$ below $T\sim 0.35$ MeV, and this is again consistent with the vanishing heat capacity as shown in Fig.~\ref{fig:cv}.
With the temperature $T\ge 0.35$ MeV, the contribution of the $s=0$ states fall down, while the contribution of the $s=2$ states go up. This corresponds to the first inflexion point in the heat capacity curve. 
Above $T\sim 0.6$ MeV, the $s=4$ states start to contribute and its contribution keeps increasing. 
Note that at $T\sim 0.8$ MeV, the contribution from the high energy $s=2$ (one pair broken) states starts to be extremely suppressed, and thus the increase of corresponding seniority component becomes slower. This is just the reason for the S shape of the heat capacity curve. It might be possible that at higher temperature, where the high energy $s=4$ (two pair broken) states are suppressed, more inflexions would appear and a second S shape would be shown.
Apart from these states, the contribution of all the other states with $s\ge 6$  is negligible.
Furthermore, it is worthwhile to mention that only part of the Cooper pairs are broken at the high temperature $T=1$ MeV, and this is just the reason for the nonvanishing pairing gap energies at high temperature as shown in Fig.~\ref{fig:gap}.



\begin{figure}[ht!]
\centering
 \includegraphics[scale=.8]{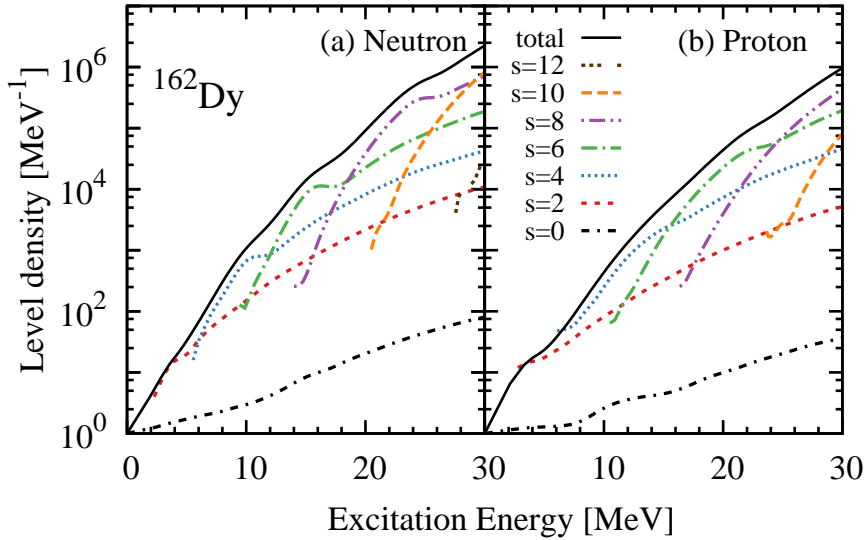}
 \caption{(Color online) Neutron (a) and proton (b) level densities for $^{162}$Dy with different seniority numbers $s = 0, 2, 4, 6, 8, 10, 12$ and the total contribution as functions of the excitation energies.}
 \label{fig:level_density}
\end{figure}

The structure of level density has been introduced to characterize pairing transition in hot nuclei. In Refs.~\cite{Dossing1995PRL75:1276--1279,Schiller2001PRC63:021306}, for instance, they considered that the level density is roughly composed of two components: (i) a low energetic part; approximately a straight line in the log plot, and (ii) a high energetic part including the theoretical Fermi gas extrapolation; a slower growing function. In our calculations, the number of the excited states of nuclei can be obtained after diagonalization of Hamiltonian (Eq.~(\ref{eq:h})) and studied in terms of different seniority contributions. In Fig.~\ref{fig:level_density}, the neutron (a) and proton (b) level densities for $^{162}$Dy with $s = 0, 2, 4, 6, 8, 10, 12$ and the total contribution as functions of the excitation energies are shown. 
It can be found that the total level density curves are smooth for both neutron and proton. The straight lines are presented at the excitation energies smaller than 4 MeV for neutron and 3 MeV for proton. Slower growing and small fluctuations are shown at higher excitation energies. However, these critical points are very inconspicuous in the level density plots. Therefore, the total level densities are decomposed into different seniority contributions as shown in Fig.~\ref{fig:level_density}. It shows that the level densities of fully paired states ($s = 0$) for both neutrons and protons increase almost linearly with the excitation energies. Meanwhile, the level density curves of $s = 2, 4, 6,$ and $8$ for neutrons present clear protuberances, as well as those curves of $s = 4$ and $6$ for protons. For higher $s$ states, protuberances do not appear. 
It can be understood now that the first fluctuation at 4 MeV for neutron and 3 MeV for proton are ascribed to the contribution of corresponding one-pair-broken states. The number of the excited states increase rapidly due to the appearance of one-pair-broken states with the increasing excitation energies. Beyond above energies, the number of states cannot increase as the same rate as before since the new one-pair-broken states cannot be formed any more. This behavior just corresponds to the first S shape of heat capacity. Similarly, when the second fluctuation comes, the two-pair-broken states appear rapidly and drop after 10 MeV excitation energy for neutron and 12 MeV for proton. Other protuberances may suggest extra S shape in heat capacity curve at higher energy. For the level density with even higher seniority, protuberances do not appear up to 30 MeV excitation energy due to the limitation of our model space. Similarly, the smooth total level density and disrupted level density by pairing are also discussed in~\cite{Dossing1995PRL75:1276--1279}.



\begin{figure}[ht!]
\centering
 \includegraphics[scale=.8]{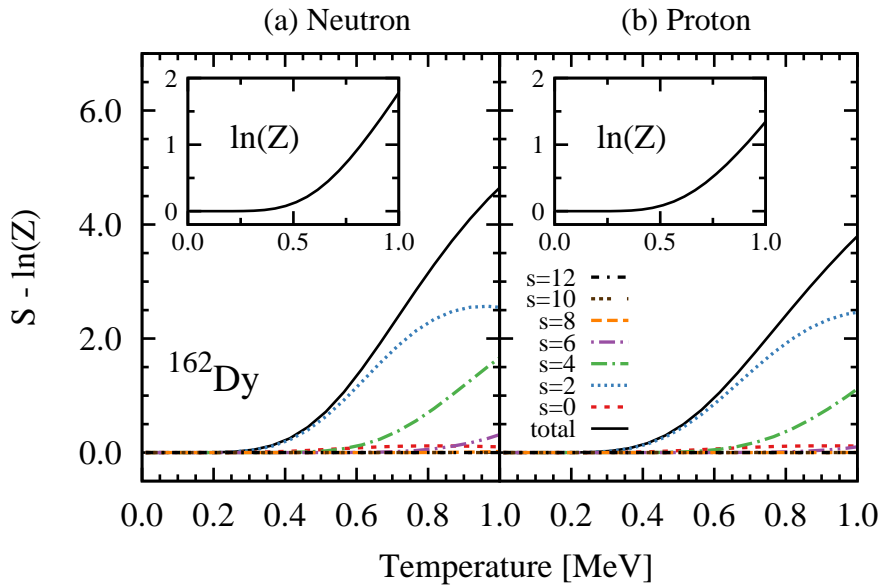}
 \caption{(Color online) Neutron (a) and proton (b) entropies subtract by $\ln Z$ for $^{162}$Dy for different seniority numbers $s = 0, 2, 4, 6, 8, 10, 12$ and the total contribution as functions of the temperature. Insert: $\ln Z$ as functions of the temperature.}
 \label{fig:entropy}
\end{figure}

With the help of defined seniority components, the thermodynamic properties of hot nuclei can be studied as well. The entropy can be evaluated by $S = \frac{\langle E(T) \rangle}{T} + \ln Z$. According to the definition of the seniority component, the average energy can be decomposed into different seniority component contributions easily. Then one can study the contributions of different seniority states to the entropy of hot nuclei. In Fig.~\ref{fig:entropy}, the neutron (a) and proton (b) entropies subtracted by $\ln Z$ for $^{162}$Dy with different seniority numbers $s = 0, 2, 4, 6, 8, 10, 12$ and the total contribution as functions of the temperature are illustrated. The curves of $\ln Z$ for neutrons and protons as functions of temperature are also shown in the insert figures. It can be found that the total entropy minus $\ln Z$ for both neutron and proton are zero at $T < 0.35$ MeV. Beyond 0.35 MeV, these curves increase almost linearly. The $s = 2$ states contribute at $T > 0.35$ MeV. The $s = 4$ states appear at $T > 0.6$ MeV. However, other higher seniority states do not contribute to the entropy up to $T = 1$ MeV.
It can be understood by the fact that the $s = 0$ states do not absorb any energy. Particles just undergo transitions among single particle levels. However, for those $s \neq 0$ states, the entropies increase with respect to temperature because they need extra energy to break particle pairs. These procedures are irreversible. Due to the limitation of model space, higher seniority contributions to the entropy are not presented up to 1 MeV.


\section{Summary}

In summary, the pairing correlations in hot nuclei $^{162}$Dy have been investigated by covariant density functional theory, and the paring correlations have been treated by a shell-model-like approach, in which the particle number is conserved exactly. 
The thermodynamical quantities have been evaluated in the canonical ensemble theory, and a clear S shape of the heat capacity curve with respect to the temperature has been presented. Due to the effect of the particle-number conservation, the pairing gap varies smoothly with the temperature, indicating a gradual transition from the superfluid to the normal state. The level density and the entropy are also investigated in terms of different seniority components.  It is found that the $s=2$ (one pair broken) states play crucial roles in the appearance of the S shape of the heat capacity curve. Higher seniority sates may lead to higher term S shape of heat capacity at high energy.

It should be noted that the present SLAP calculation has provided an exact treatment of pairing correlations without any particle number fluctuations. In the future, it will be also very interesting to investigate the odd-$A$ hot nuclei with the present framework since the blocking effects here can be treated exactly. Moreover, the SLAP could be implemented easily and effectively to the self-consistent framework of the density functional theory, to develop a self-consistent finite temperature CDFT+SLAP model would also provide significant insights in the investigation of hot nuclei. The related work are still in progress. 

\begin{acknowledgments}
The authors are grateful to Shuangquan Zhang, Neculai Sandulescu, and Jie Meng for helpful discussions. This work was partly supported by ``the Fundamental Research Funds for the Central Universities"(JUSRP1035), National Natural Science Foundation of China under Grant Nos. 11305077, 11275098 and 11505058.
\end{acknowledgments}


\bibliographystyle{apsrev4-1}
\bibliography{heat}



\end{document}